# Dynamic State Estimation of Generators under Cyber Attacks

Yang Li [1], Senior Member, IEEE, Zhi Li [1], Liang Chen [2]

[1] School of Electrical Engineering, Northeast Electric Power University, Jilin 132012, China
[2] School of Automation, Nanjing University of Information Science & Technology, Nanjing 210044, China

Corresponding author: Yang Li (e-mail: liyang@neepu.edu.cn).

This work was supported in part by the China Scholarship Council (CSC) under Grant 201608220144, the National Natural Science Foundation of China under Grant No. 51677023.

**ABSTRACT** Accurate and reliable estimation of generator's dynamic state vectors in real time are critical to the monitoring and control of power systems. A robust Cubature Kalman Filter (RCKF) based approach is proposed for dynamic state estimation (DSE) of generators under cyber attacks in this paper. First, two types of cyber attacks, namely false data injection and denial of service attacks, are modelled and thereby introduced into DSE of a generator by mixing the attack vectors with the measurement data; Second, under cyber attacks with different degrees of sophistication, the RCKF algorithm and the Cubature Kalman Filter (CKF) algorithm are adopted to the DSE, and then the two algorithms are compared and discussed. The novelty of this study lies primarily in our attempt to introduce cyber attacks into DSE of generators. The simulation results on the IEEE 9-bus system and the New England 16-machine 68-bus system verify the effectiveness and superiority of the RCKF.

## I. INTRODUCTION

In the operation of a power system, it is of critical importance to obtain accurate and reliable dynamic state vectors in real time for monitoring and controlling of a generator [1-3]. As a typical cyber-physical system, power systems have emerged and been in operation for more than a hundred years, but the increasing number of cyber attacks, natural disasters, and reliance on communication control has led to new sources of failure and cascading failure. In addition, with the large-scale integration of renewable energy and new power electronics, the uncertainty of power systems has also increased [4-6]. All these changes have brought new challenges to maintaining safe and reliable operation of the system. At the same time, the successful use of phasor measurement units (PMUs) provides new means for power system state estimation, stability assessment, and control [7, 8].

The term "dynamic state estimation" (DSE) dates back to the 1970s [9]. The Kalman filtering method was previously used to estimate the state of power systems. Recently, there have been many studies on state estimation based on electromechanical transients [10-11]. Various nonlinear filtering methods have been utilized for the DSE of generators. For example: Particle Filter (PF) [12-13], extended Kalman filter (EKF) [14-15], unscented Kalman filter (UKF) [16-17], cubature Kalman filter (CKF) [18-19]. In general, UKF is superior to EKF in terms of estimated performance; in fact, the specific results depend on the distribution of sigma points and the state dimension.

Today's power systems are facing increasingly serious cyber threats, among which false data injection (FDI) and denial of service (DoS) attacks are two typical attacks among them. The FDI attack considers measurement equipment as an attack object in power systems and has the ability to avoid bad data detection, which deviates the actual estimate from the normal value. In the operation and monitor of power systems, FDI attacks worsen online security assessment of power systems. In terms of economic benefits, FDI attacks can affect the normal dispatching plan of power systems, increase the operating cost of the system, and may even cause a large-scale blackout accident due to the wrong scheduling plan resulting from the attacks. On December 23, 2015, the large-area blackout in Ukraine was a real case of FDI attacking the power system [20]. As another typical cyber attack, the DoS attack aims to continuously transmit the forged data packets on the communication channel of cyber, so that the communication channel connecting the control center and the remote terminal is not working properly. The information can't be delivered normally, resulting in the loss of the data packet and erroneous estimation results. Therefore, these cyber attacks pose a significant threat to the normal operation power systems.

In recent years, cyber attacks have become a hot topic in power system studies. In [21], a detection model based on extreme learning machine was proposed to test and identify FDI attacks. A DSE-based risk mitigation strategy was presented for eliminating threat levels from cyber attacks in [22]. Furthermore, reference [23] developed a hybrid filtering algorithm to deal with the attacks of power systems by considering the influence of PMU. In [24], from the perspective of the attacker, the FDI attack based on the DC power flow model for state estimation was proposed. The principal component analysis method and the sparseness of the attack matrix were used to study FDI attacks [25, 26]. Moreover, in [27], the influence of Kalman filter based on model uncertainty and malicious cyber attacks on the dynamic estimation of power systems was studied. Reference [28] showed that the attacker constructs an undetectable attack vector for AC state estimation by studying in a special region of the system. In terms of detection of the FDI attack, a short-term state prediction method was proposed in [29]. For the DoS attacks, the references of [30-33] are as follows: A special DoS attack mode for the performance of a cyber-physical system was proposed in [30]. The best DoS attack plan and scheme based on the cost function was studied in [31]. Besides, in [32], the remote state estimation of cyber-physical systems under DoS attacks based on signal interference was researched. For the general system, the algorithm about attack power allocation was proposed in [33]. In summary, in the field of state estimations, the existing methods only consider the inclusion of cyber attacks in the power flow model. To the best of authors' knowledge, until now no study in the literature has reported the DSE of generators under cyber attacks.

In this paper, an RCKF-based DSE method is proposed for generators under cyber attacks. First, the FDI and DoS attacks are modelled and thereby introduced into the DSE by mixing the attack vectors with the measurement data; second, the RCKF algorithm and the CKF algorithm are adopted to the DSE under cyber attacks, and then the two algorithms are compared and discussed. The main contributions of this paper are as follows:

(1) This work is the first attempt to perform dynamic state estimation of generators under cyber attacks. Two types of cyber attacks, namely FDI and DoS, are modelled and for the first time introduced into DSE of a generator.

(2) Under cyber attacks with different degrees of sophistication, the RCKF algorithm and the CKF algorithm are adopted to the DSE of generators, and then the two algorithms are compared and discussed.

(3) The simulation results on the IEEE 9-bus system and the New England 16-machine 68-bus system verify the effectiveness and applicability of the RCKF under different attacks. Furthermore, the results also demonstrate that the RCKF performs better than the CKF in the presence of cyber attacks.

The rest of this paper is organized as follows: the generator model is described in Section II. Section III introduces the modelling of attack models. Section IV gives the detailed RCKF-based DSE of generators under cyber attacks. Section V demonstrates simulation results on two IEEE test systems, and finally, the conclusions are drawn in section VI.

## II. GENERATOR MODEL

Generally, in the process of estimating the state of power systems, the system equations and the measurement equations are concentrated in the following equation [1-3, 22, 27, 31-34]

$$\begin{cases} x_{k+1} = f(x_k, \bar{u}_k, v_k) \\ z_{k+1} = h(x_{k+1}, \bar{u}_{k+1}, w_k) \end{cases} \quad (1)$$

where $x$ represents the state vector, $\bar{u}$ refers to the control vector, and $z$ is the measurement vector; $f$ and $h$ are the state equation and the measurement equation, respectively; $v$ and $w$ are system deviation and measurement deviation respectively, which obey the normal distribution with the mean 0 and the variance matrices $Q_k$ and $R_k$. Here, $Q_k$ and $R_k$ are the system and measurement noise variance matrices, $k$ is the moment.

The fourth-order transient model of a generator is shown as follows [1, 2, 22, 27, 34]:

$$\begin{cases} \dot{\delta} = \omega - 1 \\ \dot{\omega} = \frac{1}{T_J}\left[T_m - T_e - D(\omega - 1)\right] \\ \dot{E}_q^{'} = \frac{1}{T_{d0}^{'}}\left[E_f - E_q^{'} - (X_d - X_d^{'})i_d\right] \\ \dot{E}_d^{'} = \frac{1}{T_{q0}^{'}}\left[-E_d^{'} + (X_q - X_q^{'})i_q\right] \end{cases} \quad (2)$$

where $E_d^{'}$ is the d-axis transient voltage of a generator; $E_q^{'}$ is the q-axis transient voltage of a generator; $\omega$ is the rotor speed; $T_J$ is the inertia time constant; $T_m$ represents the mechanical torque. $E_f$ is the field voltage; $T_e$ represents the electromagnetic torque; $D$ represents the damping coefficient; $\delta$ is the rotor angle. $T_{d0}^{'}$ and $i_d$ are respectively d-axis transient time constant and d-axis output currents; $X_d$ and $X_d^{'}$ are d-axis reactance and d-axis transient reactance; $T_{q0}^{'}$ and $i_q$ are respectively q-axis transient time constant and q-axis output currents, respectively; $X_q$ and $X_q^{'}$ are respectively q-axis reactance and q-axis transient reactance.

Measurement vectors include $\delta$, $\omega$ and $X_q$. The measurement equation is listed as follows [1, 3]:

$$\begin{cases} \delta^z = \delta \\ \omega^z = \omega \\ P_e^z = \dfrac{U^2}{2}\sin(2\delta - 2\varphi)\left(\dfrac{1}{X_q'} - \dfrac{1}{X_d'}\right) \\ \quad + \dfrac{U\sin(\delta-\varphi)E_q'}{X_d'} + \dfrac{U\cos(\delta-\varphi)E_d'}{X_q'} \end{cases} \quad (3)$$

where $\omega^z$ and $\delta^z$ are the measurements of the rotor speed and the rotor angle, respectively. $P_e^z$ denotes the measurements of the electromagnetic power of a generator. $U$ and $\varphi$ are the magnitude and phase angle of the generator terminal voltage [1].

The measurement noise covariance matrix $R_{k+1}$ is

$$R_{k+1} = \begin{bmatrix} \sigma_{\delta_z}^2 & 0 & 0 \\ 0 & \sigma_{\omega_z}^2 & 0 \\ 0 & 0 & \sigma_{P_{e_z}}^2 \end{bmatrix} \quad (4)$$

where $\sigma_{\delta_z}^2$ and $\sigma_{\omega_z}^2$ are the measurement variance of rotor angle and rotor speed, and $\sigma_{P_{e_z}}^2$ is the measurement variance of electromagnetic power.

$$P_{e_z}^2 = \left(\dfrac{\partial P_e}{\partial U}\right)^2 \sigma_U^2 + \left(\dfrac{\partial P_e}{\partial \varphi}\right)^2 \sigma_\varphi^2 \quad (5)$$

where $\sigma_U = 0.2\%$, $\sigma_\varphi = 0.2°$.

In the DSE of a generator, the state vector, the measurement vector, and the control vector are specifically shown as [1, 29]:

$$\begin{aligned} x &= \left[\delta, \omega, E_q', E_d'\right]^T \\ z &= \left[\delta^z, \omega^z, P_e^z\right]^T \\ \bar{u} &= \left[T_m, E_f, U, \varphi\right]^T \end{aligned} \quad (6)$$

The mechanical torque and the field voltage can be obtained from the governor model and the exciter model respectively. The governor model is shown below:

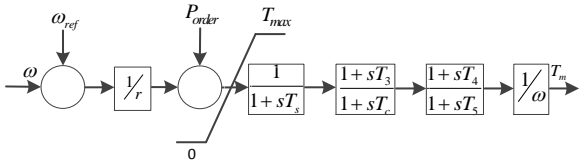

**FIGURE 1.** Governor model

where $\omega_{ref}$ is the rated speed of the generator rotor, $1/r$ is the steady-state gain, $T_{\max}$ is the maximum power order, $T_s$ is the servo time constant, $T_c$ refers to the turbine time constant, $T_3$ is the transient gain time constant, $T_4$ is the time constant to set percentage, $T_5$ is the reheater time constant.

The exciter model is shown as follows:

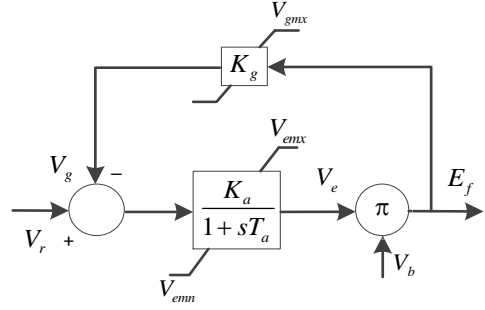

**FIGURE 2.** Exciter model

where $V_r$ is the regulator output voltage, $K_a$ is the voltage regulator gain, $T_a$ is the voltage regulator time constant, $V_b$ is the potential circuit voltage output, $K_g$ is the feedback constant, $E_f$ is the field voltage.

### III. ATTACK MODELS

#### A. FDI ATTACK

For the DSE of generators, state vectors are uniformly represented as $x = [x_1, x_2, x_3, \ldots x_k]^T$. The measurement vectors are uniformly expressed as $z = [z_1, z_2, z_3, \ldots, z_l]^T$. The measurement equation is as follows:

$$z = h(x) + e \quad (7)$$

where

$$\begin{aligned} h(x) = [&h_1(x_1, \ldots, x_k), \\ &h_2(x_1, \ldots, x_k), \\ &\ldots \\ &h_l(x_1, \ldots, x_k)]^T \end{aligned} \quad (8)$$

represents the nonlinear relationship between the state vector and the measurement vector, and $e$ denotes the measurement error.

Take the function $h_1()$ as an example, Taylor's formula is used to extend $h_1()$. Since the state vectors of generators can't change suddenly during the electromechanical transient process, the high-order parts of $h_1()$ are too small and can be ignored. Therefore, the linearized measurement equation is obtained as follows:

$$\begin{aligned} z_1 &= h_1(x_1, \ldots, x_k) + e_1 \\ &= h_1(x_1^{(0)}, \ldots, x_k^{(0)}) + \dfrac{\partial h_1}{\partial x_1}\bigg|_0 (x_1 - x_1^{(0)}) \\ &\quad + \ldots + \dfrac{\partial h_1}{\partial x_k}\bigg|_0 (x_k - x_k^{(0)}) + e_1 \\ &= h_1(x_1^{(0)}, \ldots, x_k^{(0)}) + \left(\dfrac{\partial h_1}{\partial x_1}\bigg|_0 x_1 - \dfrac{\partial h_1}{\partial x_1}\bigg|_0 x_1^{(0)}\right) \\ &\quad + \ldots + \left(\dfrac{\partial h_1}{\partial x_k}\bigg|_0 x_k - \dfrac{\partial h_1}{\partial x_k}\bigg|_0 x_k^{(0)}\right) + e_1 \\ &= \dfrac{\partial h_1}{\partial x_1}\bigg|_0 x_1 + \ldots + \dfrac{\partial h_1}{\partial x_k}\bigg|_0 x_k + \bar{e}_1 \end{aligned} \quad (9)$$

where $x_1^{(0)},...,x_k^{(0)}$ are initial values of $x_1,...,x_k$, and they are close to $x_1,...,x_k$; $\left.\frac{\partial h_1}{\partial x_1}\right|_0,...,\left.\frac{\partial h_1}{\partial x_k}\right|_0$ are the calculation results by introducing $x_1^{(0)},...,x_k^{(0)}$ into these partial derivatives; $\overline{e}_1$ is a constant matrix. Thereby, (7) can be further expressed as follows:

$$z = Hx + \overline{e} \quad (10)$$

$$H = \begin{bmatrix} \frac{\partial h_1}{\partial x_1} & \frac{\partial h_1}{\partial x_2} & \frac{\partial h_1}{\partial x_3} & \cdots & \frac{\partial h_1}{\partial x_k} \\ \frac{\partial h_2}{\partial x_1} & \frac{\partial h_2}{\partial x_2} & \frac{\partial h_2}{\partial x_3} & \cdots & \frac{\partial h_2}{\partial x_k} \\ & & \cdots & & \\ \frac{\partial h_l}{\partial x_1} & \frac{\partial h_l}{\partial x_2} & \frac{\partial h_l}{\partial x_3} & \cdots & \frac{\partial h_l}{\partial x_k} \end{bmatrix} \quad (11)$$

where $H$ is the Jacobian matrix, $\overline{e}$ is regarded as the new measurement error. In this study, by substituting (3) into (11), the matrix $H$ can be obtained as

$$H = \begin{bmatrix} 1 & 0 & 0 & 0 \\ 0 & 1 & 0 & 0 \\ L_1 & 0 & L_2 & L_3 \end{bmatrix} \quad (12)$$

where $L_1 = \frac{\partial P_e^z}{\partial \delta}$, $L_2 = \frac{\partial P_e^z}{\partial E_q'}$, $L_3 = \frac{\partial P_e^z}{\partial E_d'}$.

Bad data that often occurs with a large terrible impact on the results of state estimation The application of the residual equation can largely eliminate the influence of bad data [29].

The residual between $z$ and its estimated vector is given by the following equation [28]:

$$r = z - \hat{z} \quad (13)$$

where $\hat{z} = h(\hat{x})$ represents the estimated vector of $z$.

According to the definition of the residual, Eq. (13) can be rewritten as follows [20, 21, 24, 25]:

$$r = z - H\hat{x} \quad (14)$$

If $a = [a_1,...,a_l]^T$ is used to represent the false data injected by the attacker in the measurement vector, the actual measurement vector is $z_a = z + a$. $c = [c_1,...,c_k]^T$ represents the error vector that is brought into by the FDI attack in the state vector, the elements of it are randomly generated by a Gaussian random variable with different variances, and the estimation of state vector becomes $\hat{x}_a = \hat{x} + c$ [25]. According to (14), the measurement residual is obtained as [15]

$$\begin{aligned} \|r\| &= \|z_a - H\hat{x}_a\| \\ &= \|z + a - H(\hat{x} + c)\| \\ &= \|z - H\hat{x} + a - Hc\| \end{aligned} \quad (15)$$

Obviously, $a = Hc$ is a sufficient condition for (16).

$$\|r\| = \|z_a - H\hat{x}_a\| = \|z - H\hat{x}\| \quad (16)$$

Eq. (16) suggests that the residual values before and after the FDI attack are equal, and then the residual-based bad data detection is unable to identify the false data. Consequently, the FDI attack is successfully applied to the measurement vector when the attack is modelled as $a = Hc$. In this case, the measurement vector under the attack has a larger deviation from the true vector, which will undermine the safe and stable operation of generators [28].

If the errors of attack vectors are taken into account, the residual values before and after the FDI attack are not equal [29]. Then the following formula is obtained:

$$\begin{aligned} \|z_a - H\hat{x}_a\| &= \|z - H\hat{x} + a - Hc\| \\ &\leq \|z - H\hat{x}\| + \|a - Hc\| \end{aligned} \quad (17)$$

However, if the residual value of the measurement data is less than the detection threshold in the detection process, the FDI attacks are still successfully hidden. Furthermore, the detection threshold $B_J$ is determined by superimposing a certain redundancy on the normal maximum estimated deviation. The formula of the detection is as follows:

$$\|z - H\hat{x}\| \leq \|z_a - H\hat{x}_a\| \leq B_J \quad (18)$$

Thus, if $z_a$ satisfies (18), the FDI attacks can be implemented to the DSE of generators.

### B. DoS ATTACK

The essence of DoS attacks is the process of packet (i.e., measurement data) loss. At present, there are usually two types of modelling for the characteristic of packet loss. The first one is described by the Bernoulli distribution [31], while the second is described by the Markov model [33]. Taking into account the characteristics of the memoryless Bernoulli process in DoS attacks, the first method is here chosen to model DoS attacks.

When an attacker initiates $d$ consecutive DoS attacks, it may result in successive loss of packets. For example, while the measurement data at the ($k$-$d$)th moment is successfully transmitted, the attacker launches an attack at the next moment. The period of consecutive attacks is from the ($k$-$d$+1)th moment to the $k$th moment, and up to $d$ packets are lost during this period. Then, the Bernoulli distribution is used to depict the packet loss characteristics caused by DoS attacks. In order to describe the transmission of the measurement data $z_k, z_{k-1},...,z_{k-d+1}, z_{k-d}$. Define the row matrix $\mu_k \in \mathbf{R}^{1 \times (d+1)}$, the elements in this matrix can be expressed as follows [30-33]:

$$\mu_k(i) = \begin{cases} 1 \\ 0 \end{cases} \quad (19)$$

$$\begin{aligned} var(\mu_k(i) = 1) &= P_s(\mu_k(i) = 0) \cdot P_s(\mu_k(i) = 1) \\ &\triangleq \rho \cdot (1 - \rho) \end{aligned} \quad (20)$$

where $\mu_k$ is the state matrix of measurement vector transmission, $\mu_k(i)$ denotes the element $i$ in the matrix $\mu_k$ ($i \in [1, d+1]$), representing the transmission state of the measurement $z_{k-i+1}$ at the ($k$-$i$+1)th moment. If $\mu_k(i) = 0$, the measurement $z_{k-i+1}$ is lost; otherwise, it is successfully transmitted. $\rho \in (0,1)$ denotes the probability of packet loss. And thereby, the measurement data actually received by the state estimator under DoS attacks can be indicated as:

$$z_k' = \begin{bmatrix} \mu_k(1) \times z_k \\ \mu_k(2) \times z_{k-1} \\ \cdots \\ \mu_k(d+1) \times z_{k-d} \end{bmatrix} \quad (21)$$

where $z_k'$ denotes the measurement data under the DoS attack.

## IV. RCKF-BASED DSE OF GENERATORS

### A. CUBATURE KALMAN FILTER

The spherical cubature and the Gaussian quadrature rules are utilized to estimate the probability density functions of the state space and the measurement space in CKF. Fig. 3 shows the basic structure of the DSE for generators based on the CKF. The specific process includes the following two stages [1]:

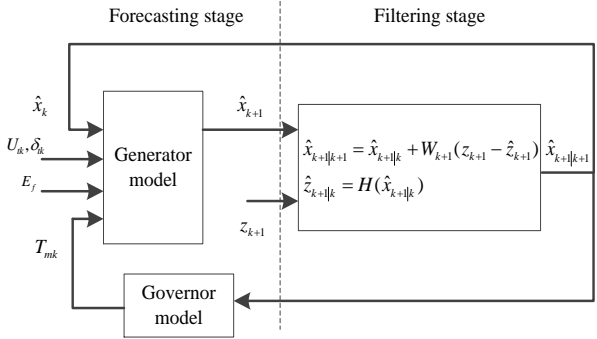

**FIGURE 3.** DSE based on CKF

(1) Forecasting stage

At this stage, $U$ and $\varphi$ are all available from PMUs. Decouple the generator from the system, and then obtain the predicted values of state vectors at moment $k+1$ by the state equation and the state vectors of at moment $k$, which is shown as follows:

$$P_{k|k} = S_{k|k} S_{k|k}^T \quad (22)$$

$$X_{i,k|k} = S_{k|k}\xi_i + \hat{x}_{k|k} \quad (23)$$

$$X_{i,k|k}^* = f\left(X_{i,k|k}, u_k\right) \quad (24)$$

$$\hat{x}_{k+1|k} = \frac{1}{2n}\sum_{i=1}^{2n} X_{i,k+1|k}^* \quad (25)$$

$$P_{k+1|k} = \frac{1}{2n}\sum_{i=1}^{2n} X_{i,k+1|k}^* X_{i,k+1|k}^{*\mathrm{T}} - \hat{x}_{k+1|k}\hat{x}_{k+1|k}^T + Q_k \quad (26)$$

where $P_{k|k}$ denotes the estimation error variance matrix of the state vector at moment $k$ obtained from moment $k-1$; $X_{i,k|k}$ represents the cubature points of the state vectors at moment $k$; $\xi_i = \sqrt{n}$, $i=1,2,...2n$, $n$ is the dimension of the state vector. $X_{i,k|k}^*$ represents the predicted values of the cubature points obtained by the state equation; $\hat{x}_{k+1|k}$ is the predicted values of the state vectors; $P_{k+1|k}$ refers to the predicted error variance matrix of the state vectors.

(2) Filtering stage

The predicted values of the state vectors at the forecasting stage are calculated by the measurement vector $z_{k+1}$, and thereafter obtain the estimation of the state vectors at moment $k+1$. The specific process is as follows:

$$P_{k+1|k} = S_{k+1|k} S_{k+1|k}^T \quad (27)$$

$$X_{i,k+1|k} = S_{k+1|k}\xi_i + \hat{x}_{k+1|k} \quad (28)$$

$$Z_{i,k+1|k} = h\left(X_{i,k+1|k}, u_k\right) \quad (29)$$

$$\hat{z}_{k+1|k} = \frac{1}{2n}\sum_{i=1}^{2n} Z_{i,k+1|k} \quad (30)$$

$$P_{zz,k+1|k} = \frac{1}{2n}\sum_{i=1}^{2n} Z_{i,k+1|k} Z_{i,k+1|k}^T - \hat{z}_{k+1|k}\hat{z}_{k+1|k}^T + R_{k+1} \quad (31)$$

$$P_{xz,k+1|k} = \frac{1}{2n}\sum_{i=1}^{2n} X_{i,k+1|k} Z_{i,k+1|k}^T - \hat{x}_{k+1|k}\hat{z}_{k+1|k}^T \quad (32)$$

$$W_{k+1} = P_{xz,k+1|k} P_{zz,k+1|k}^{-1} \quad (33)$$

$$\hat{x}_{k+1|k+1} = \hat{x}_{k+1|k} + W_{k+1}\left(z_{k+1} - \hat{z}_{k+1|k}\right) \quad (34)$$

$$P_{k+1|k+1} = P_{k+1|k} - W_{k+1} P_{zz,k+1|k} W_{k+1}^T \quad (35)$$

where $X_{i,k+1|k}$ is the cubature points of the predicted values $\hat{x}_{k+1|k}$; $Z_{i,k+1|k}$ is the cubature points of forecast values of the measurements; $\hat{z}_{k+1|k}$ is the forecast values of the measurements obtained by the weighted summation of $Z_{i,k+1|k}$; $P_{zz,k+1|k}$ is the measurement error variance matrix; $P_{xz,k+1|k}$ is the cross-error variance matrix; $W_{k+1}$ is the filter gain; $\hat{x}_{k+1|k+1}$ is the final estimation at moment $k+1$; $P_{k+1|k+1}$ is the updated estimation error variance matrix of state vectors at moment $k+1$.

### B. ROBUST CUBATURE KALMAN FILTER

Accurate system models and noise statistics are prerequisites for the traditional CKF for obtaining a good estimate. In this paper, the generator model is assumed to be accurate. However, due to environmental factors, bad data inevitably appears in PMU measurement vectors, causing the measurement error variance matrix $R_{k+1}$ to be inconsistent with the actual error. This leads to that the CKF is unable to complete the accurate correction of the predicted values in the filtering stage.

By combining the robust M estimation theory and the CKF, the RCKF has the on-line adjustment capability for measuring noise statistics. By using the RCKF, accurate state estimation results can still be obtained even if the measurements contain bad data.

The basic process of the RCKF is generally the same as that of the CKF except for (31) and (32). Specifically speaking, the measurement error variance matrix before correction is replaced by the corrected measurement error variance matrix in (31).

$R_{k+1}$ is the measurement error variance matrix before correction, $\bar{R}_{k+1}$ is the corrected measurement error variance matrix according to the following formula:

$$\bar{R}_{k+1} = \bar{P}^{-1} \quad (36)$$

where $\bar{P}$ is the equivalence weight matrix.

In this work, the Huber method is used to calculate the

equivalence weight matrix $\bar{P}$ [1], which is given by

$$\bar{p}_{m,m} = \begin{cases} \dfrac{1}{\sigma_{m,m}}, & (\left|\dfrac{r_m}{\sigma_{rm}}\right| = |r_m'| \leq C) \\ \dfrac{C}{\sigma_{m,m}|r_{m,m}'|}, & (|r_m'| > C) \end{cases} \quad (37)$$

$$\bar{p}_{m,n} = \begin{cases} \dfrac{1}{\sigma_{m,n}}, & (|r_m'| \leq C, |r_n'| \leq C) \\ \dfrac{C}{\sigma_{m,n}\max(|r_m'|,|r_n'|)}, & (|r_m'| > C, |r_n'| > C) \end{cases} \quad (38)$$

where $\bar{p}_{m,m}$ and $\bar{p}_{m,n}$ are respectively a diagonal and non-diagonal element of the matrix $\bar{P}$. In the matrix $R_{k+1}$, $\sigma_{m,m}$ and $\sigma_{m,n}$ denote a diagonal and non-diagonal element. Since the measurement error variance matrix in the DSE of generators is a diagonal matrix, the non-diagonal elements $\sigma_{m,n}$ are taken as zero. $r_m$ is the corresponding residual component of measurement vectors $z_m$, while $r_m'$ is the corresponding standard residual component. $\sigma_m$ is the mean square error of $r_m$. $C$ is a given constant (ranging from 1.3 to 2.0), and it is here taken as 1.5 through the try-and-error method. These vectors are expressed as [1]:

$$r_m' = r_m / \sigma_{rm} \quad (39)$$

$$r_m = (z_{k+1} - \hat{z}_{k+1|k})_m \quad (40)$$

$$\sigma_{rm} = (P_{zz,k+1|k})_{m,m} \quad (41)$$

where $P_{zz,k+1|k}$ is the measurement error variance matrix before correction.

In the process of the RCKF-based DSE of generators, $\hat{x}_{k+1|k}$ and $\hat{x}_{k+1|k+1}$ are approximately equal before attacks. When attack vectors are implemented into the measurement vectors, $\hat{x}_{k+1|k+1}$ obtained in the filtering stage will change. But at this time, $\hat{x}_{k+1|k}$ obtained in the forecasting stage remains unchanged. In this regard, this paper proposes an attack identification strategy, which is as follows:

$$\begin{cases} \|\hat{x}_{k+1|k+1} - \hat{x}_{k+1|k}\| > D_J, \text{The attack can be identified} \\ \|\hat{x}_{k+1|k+1} - \hat{x}_{k+1|k}\| \leq D_J, \text{The attack cannot be identified} \end{cases} \quad (42)$$

where $D_J$ is $\max\|\hat{x}_{k+1|k+1} - \hat{x}_{k+1|k}\|$ during the period of the normal estimation.

### C. SOLUTION PROCESS

As shown in Fig. 4, the RCKF-based solution process is described as follows:

1) Construction of the state and measurement equations: Based on the estimation vector $\hat{x}_k$ at moment $k$, the state equation of a generator is constructed. Assuming that the magnitude and phase angle of the generator terminal voltage $(U, \varphi)$ are available from PMUs, the measurement equation of a generator is thereby constructed.

2) Modelling of cyber attacks: The Jacobian matrix $H$ is obtained from (14). For the FDI attack, the attack vector is obtained through multiplying the Jacobian matrix by the error vector of state vector obeying Gaussian distributions with different standard deviations. (2) Regarding the DoS attack, the transmission state matrix $\mu_k$ obeying the Bernoulli distribution is established.

3) Implementation of cyber attacks: On the basis of the predicted values of state vectors, the FDI attack vectors are implemented to the measurement vectors $z_{k+1}$, and then they are detected via the bad data detection. For the DoS attack, the measurement vector under the attack is formed by multiplying the elements of the matrix $\mu_k$ by the corresponding elements of the measurement matrix.

4) Estimation results: In the filtering stage of the RCKF, the estimated values at moment $k+1$ under attacks are obtained according to (27) ~ (35). And thereby, based on the estimated values, the mechanical torque is calculated by the governor. At the same time, the estimation results need to be tested for the attack identification. After that, perform the DSE based on the RCKF until the simulation is finished.

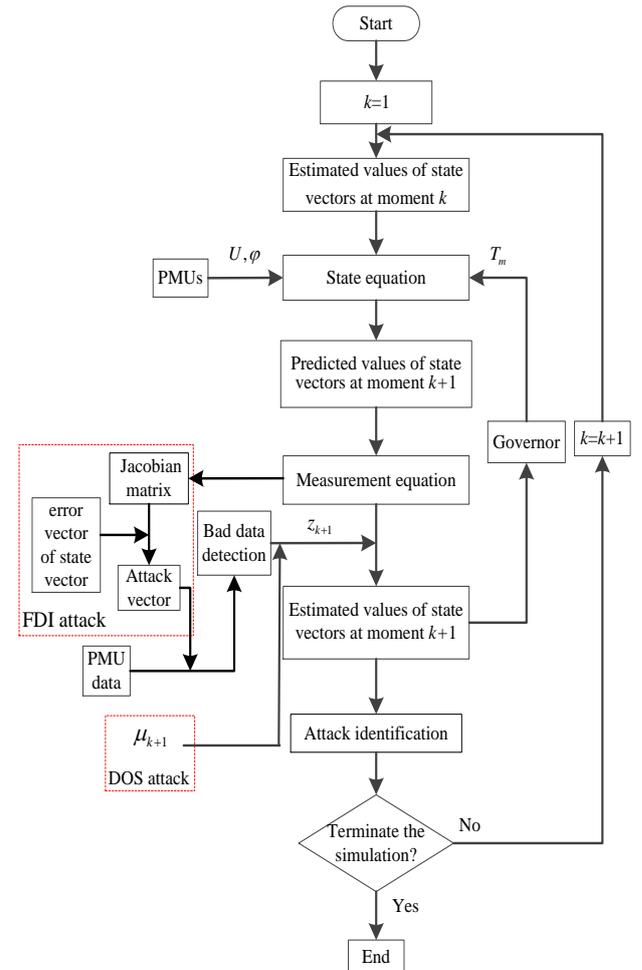

**FIGURE 4.** Solution process

## V. NUMERICAL RESULTS

### A. SIMULATION SETTINGS

All simulations have been performed under the MATLAB environment on a desktop PC equipped with Intel Core i5-4590 3.3GHz CPU and 8 GB RAM. Note that in this study, PMU data are simulated through detailed numerical simulations via the power system toolbox (PST).

The simulation data are listed as follows: 1) The sampling rate is assumed 50 samples/s; 2) The simulation time step is set to 0.02 s; 3) The standard deviations of the rotor angle and the rotor speed are respectively 2° and 0.1%; 4) The standard deviations of the phase angle and amplitude of terminal voltages are 0.1° and 0.1%; 5) A PMU is equipped at the terminal of each generator [1, 27].

### B. EVALUATION INDEX

To compare the estimation results, three different indices are defined as follows [1], [2]:

$$\tau_1 = \sqrt{\frac{1}{N}\sum_{i=1}^{N}(\frac{\hat{x}_i - x_{it}}{x_{iz}})^2} \quad (43)$$

$$\tau_2 = \sqrt{\frac{\sum_{i=1}^{N}(\hat{x}_i - x_{it})^2}{\sum_{i=1}^{N}(x_{iz} - x_{it})^2}} \quad (44)$$

$$\tau_3 = \sqrt{\frac{\sum_{i=1}^{N}(\hat{x}_i - x_{it})^2}{N}} \quad (45)$$

where $\hat{x}_i$ and $x_{it}$ are respectively the estimated value and the true value, $x_{iz}$ represents the measurement value, $N$ denotes the number of sampling points. The evaluation index $\tau_1$ can measure the filtering performance of the same state estimation method under different cyber attacks; $\tau_2$ can evaluate the filtering performances of different state estimation methods under the same cyber attack; $\tau_3$ can quantitatively evaluate the estimation results for any estimation method under any attack.

### C. IEEE 9-BUS SYSTEM

This system contains 3 generators, 3 transformers, and 9 buses. The fault settings are as follows: a three-phase short-circuit fault occurs at bus 4 at $t$=1.2s, then the fault is cleared at $t$=1.5s. The entire simulation process lasts 20s. For ease of analysis without loss of generality, generator 1 is taken as an example to examine the performances of the RCKF algorithm. Here, the detection threshold of the bad data detection is set to $B_J = 2.1$ through simulations, since all injected false data can just pass the detection while bad data can be detected in this situation.

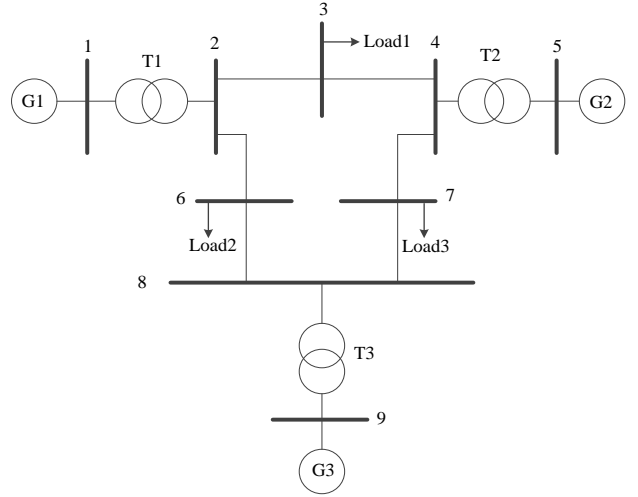

FIGURE 5. IEEE 9-bus system

1) SIMULATION RESULTS UNDER FDI ATTACKS

In terms of FDI attacks, according to the error vector of the state vector, three attack scenarios are designed, as shown in Table I [25].

TABLE I
SCENARIOS of FDI ATTACKS

| Attack Scenarios | Error Vector of State Vector |
|---|---|
| FDI-scenario 1 | $c \sim N(0, \sigma^2), \sigma = 0.0001$ |
| FDI-scenario 2 | $c \sim N(0, \sigma^2), \sigma = 0.001$ |
| FDI-scenario 3 | $c \sim N(0, \sigma^2), \sigma = 0.01$ |

In Table I, FDI-scenarios 1-3 respectively indicate that in three attack scenarios, the error vectors of the state vectors obey the Gaussian distribution with the mean 0 and the standard deviations 0.0001, 0.001 and 0.01.

Figs. 6, 8 and 10 show the estimated results of the generator's rotor angle under three FDI attack scenarios, while Figs. 7, 9 and 11 illustrate the estimation results of the rotor speed under the scenarios.

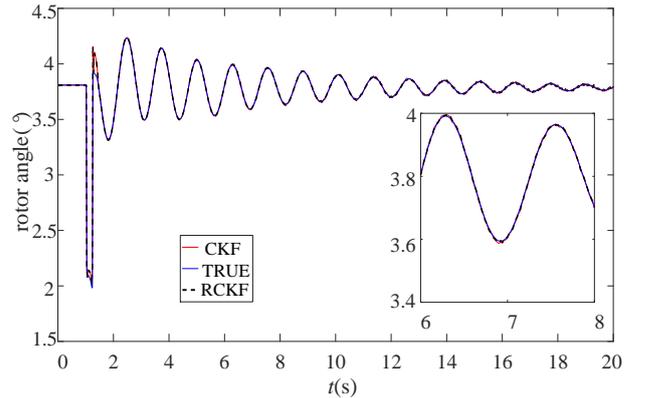

FIGURE 6. Rotor angle under FDI-scenario 1

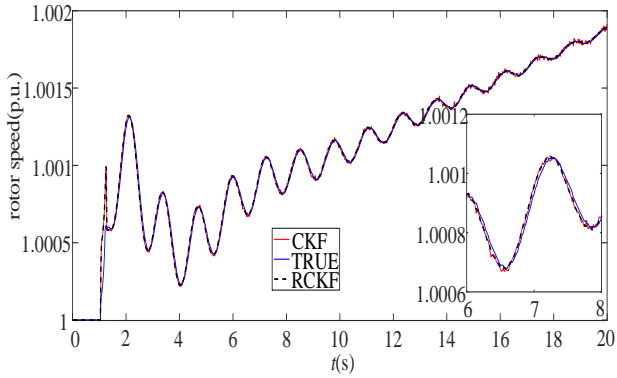

**FIGURE 7.** Rotor speed under FDI-scenario 1

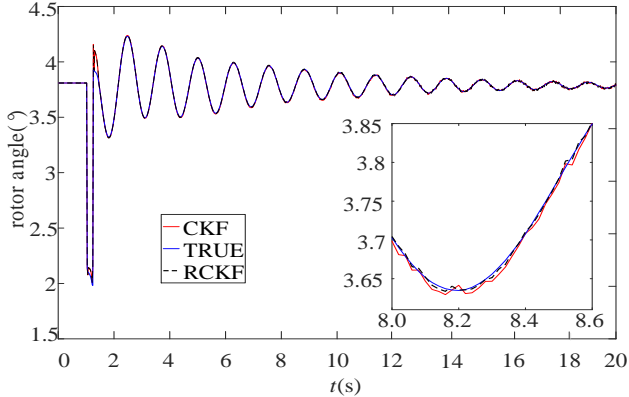

**FIGURE 8.** Rotor angle under FDI-scenario 2

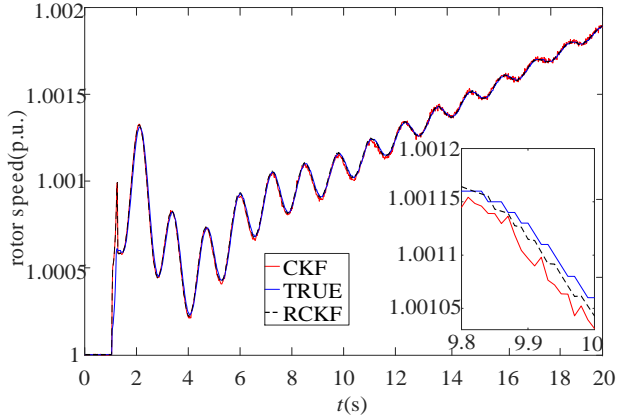

**FIGURE 9.** Rotor speed under FDI-scenario 2

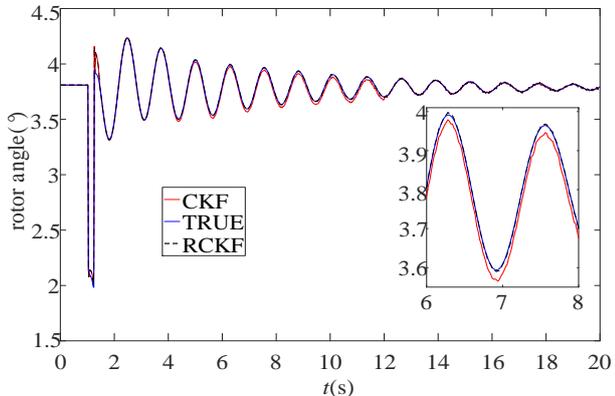

**FIGURE 10.** Rotor angle under FDI-scenario 3

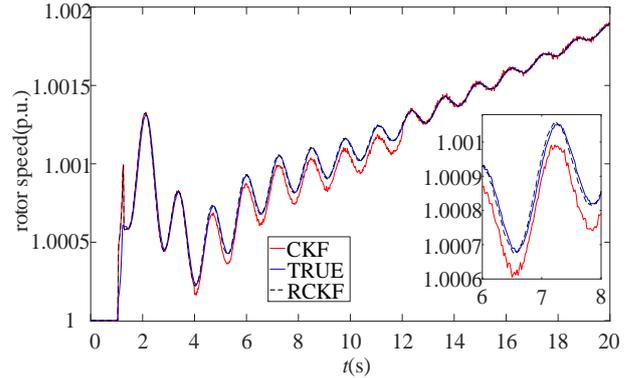

**FIGURE 11.** Rotor speed under FDI-scenario 3

From Figs. 6-11, it can be observed that the estimation results of the RCKF are very close to the true values in the attack period from $t=4$s to 12s. This result verifies the effectiveness of the RCKF under FDI attacks. Furthermore, the estimated results of the RCKF are closer to the true values than those of the CKF, which suggests that the filtering ability of the RCKF is better than that of the CKF.

The estimation indices of generator 1 in the scenarios are shown in Table II.

Table II
ESTIMATION INDICES

| Attack scenario | Index | Parameter | CKF | RCKF |
|---|---|---|---|---|
| FDI-scenario 1 | $\tau_1$ | $\delta$ | 7.8923e-04 | 6.8549e-04 |
| | | $\omega$ | 2.8102e-04 | 2.6038e-04 |
| | $\tau_2$ | $\delta$ | 0.0021 | 0.0017 |
| | | $\omega$ | 1.7812e-05 | 1.6503e-05 |
| | $\tau_3$ | $\delta$ | 7.0630e-05 | 6.0998e-05 |
| | | $\omega$ | 1.7014e-05 | 1.5821e-05 |
| FDI-scenario 2 | $\tau_1$ | $\delta$ | 8.9725e-04 | 6.8921e-04 |
| | | $\omega$ | 3.2047e-04 | 2.6147e-04 |
| | $\tau_2$ | $\delta$ | 0.0025 | 0.0018 |
| | | $\omega$ | 2.2344e-05 | 1.6386e-05 |
| | $\tau_3$ | $\delta$ | 7.9030e-05 | 6.4476e-05 |
| | | $\omega$ | 2.1003e-05 | 1.5841e-05 |
| FDI-scenario 3 | $\tau_1$ | $\delta$ | 0.0091 | 6.9046e-04 |
| | | $\omega$ | 0.0019 | 2.6201e-04 |
| | $\tau_2$ | $\delta$ | 0.0232 | 0.0019 |
| | | $\omega$ | 1.0908e-04 | 1.7344e-05 |
| | $\tau_3$ | $\delta$ | 6.6115e-04 | 6.8111e-05 |
| | | $\omega$ | 1.0408e-04 | 1.5969e-05 |

Table II gives the state estimation indices of generator 1 based on the CKF and the RCKF under three FDI attack scenarios. From Table II, it can be seen that: (1) For the index $\tau_1$, the values based on the RCKF are not much different from each other. (2) As far as index $\tau_2$ is concerned, under attack scenario 1, the filtering accuracies of the RCKF are respectively 19% and 7.3% higher than those of the CKF for the rotor angle and the rotor speed; under attack scenario 2, the filtering accuracies of the RCKF are increased by 28% and 27% compared with those of the CKF; under attack scenario 3, the filtering accuracies of the RCKF are increased by 91.8% and 84% compared with those of the CKF. (3) Regarding the index $\tau_3$, the index values of the RCKF are less than those of the CKF. In general, the effectiveness of the RCKF is verified under different FDI attack scenarios. The superiority of the RCKF under these scenarios compared with the CKF is also confirmed.

2) SIMULATION RESULTS UNDER DOS ATTACKS

According to the packet loss rate of DoS attacks, four attack scenarios are designed, as shown in Table III [31].

Table III
SCENARIOS of DoS ATTACKS

| Attack Scenarios | Packet Loss Rate |
| --- | --- |
| DoS-scenario 1 | 1.00 |
| DoS-scenario 2 | 0.95 |
| DoS-scenario 3 | 0.85 |
| DoS-scenario 4 | 0.75 |

In Table III, DoS-scenarios 1-4 respectively indicate that in four attack scenarios, the packet loss rate of the measurement data is 1, 0.95, 0.85 and 0.75.

Figs. 12, 14, 16 and 18 show the estimated results of the generator's rotor angle under four DoS attack scenarios. Figs. 13, 15, 17 and 19 show the estimated results of rotor speed under these DoS attack scenarios.

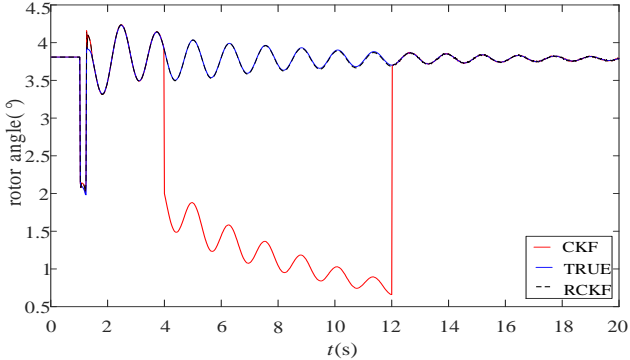

FIGURE 12. Rotor angle under DoS-scenario 1

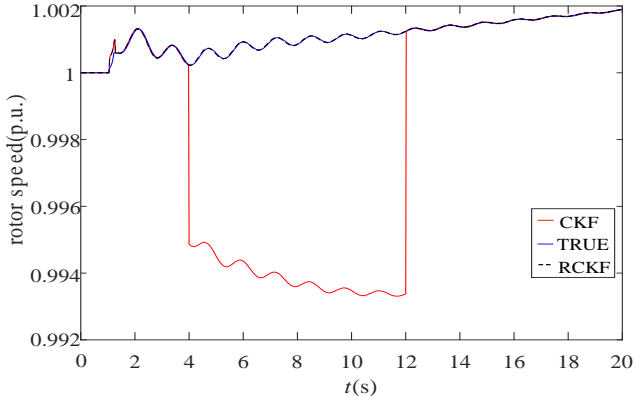

FIGURE 13. Rotor speed under DoS-scenario 1

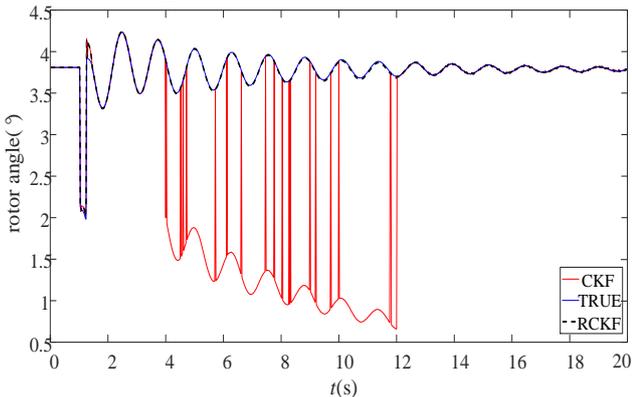

FIGURE 14. Rotor angle under DoS-scenario 2

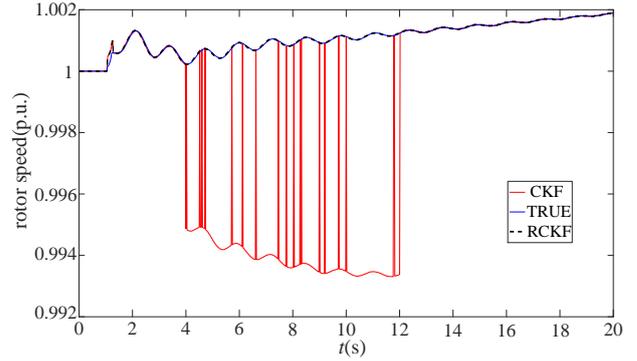

FIGURE 15. Rotor speed under DoS-scenario 2

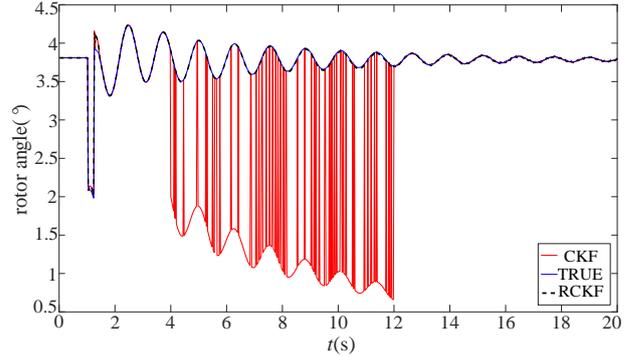

FIGURE 16. Rotor angle under DoS-scenario 3

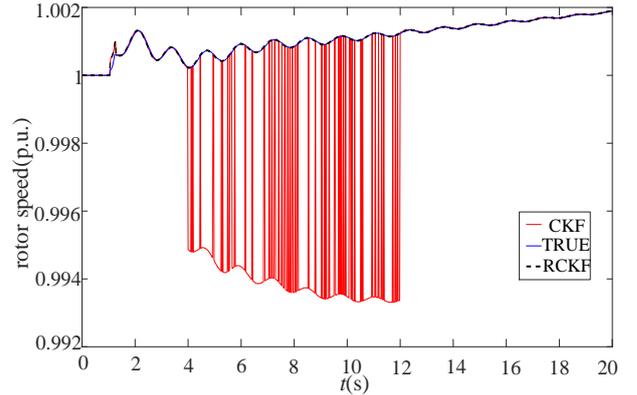

FIGURE 17. Rotor speed under DoS-scenario 3

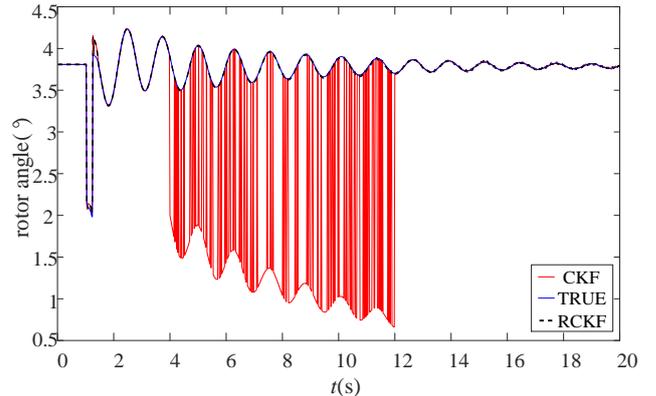

FIGURE 18. Rotor angle under DoS-scenario 4

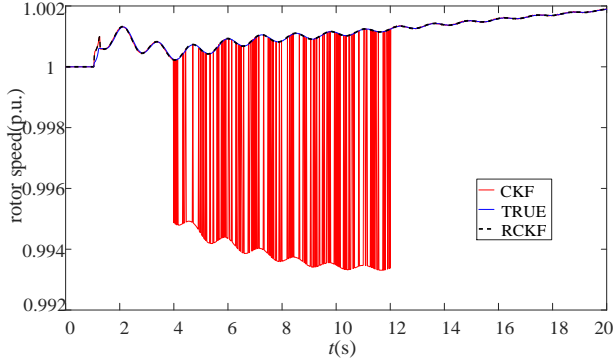

**FIGURE 19.** Rotor speed under DoS-scenario 4

From Figs. 12-19, ones can observe that in the attack period from $t$=4s to 12s, the attack frequency changes with the decrease of the packet loss rate, and the estimated values of the RCKF are much closer to the true values than those of the CKF. Based on these results, the effectiveness and superiority of the RCKF are verified under DoS attacks.

The state estimation indices of generator 1 are shown in Table IV.

Table IV
ESTIMATION INDICES

| Attack scenario | Index | Parameter | CKF | RCKF |
|---|---|---|---|---|
| DoS-scenario 1 | $\tau_1$ | $\delta$ | 0.6145 | 0.0019 |
| | | $\omega$ | 0.1409 | 2.5930e-04 |
| | $\tau_2$ | $\delta$ | 0.0251 | 9.9827e-05 |
| | | $\omega$ | 0.0074 | 1.5696e-05 |
| | $\tau_3$ | $\delta$ | 0.0458 | 1.6683e-04 |
| | | $\omega$ | 0.0071 | 1.5626e-05 |
| DoS-scenario 2 | $\tau_1$ | $\delta$ | 0.5866 | 0.0018 |
| | | $\omega$ | 0.1347 | 2.5925e-04 |
| | $\tau_2$ | $\delta$ | 0.0253 | 9.9477e-05 |
| | | $\omega$ | 0.0070 | 1.5685e-05 |
| | $\tau_3$ | $\delta$ | 0.0448 | 1.6640e-04 |
| | | $\omega$ | 0.0069 | 1.5619e-05 |
| DoS-scenario 3 | $\tau_1$ | $\delta$ | 0.5258 | 0.0017 |
| | | $\omega$ | 0.1192 | 2.5918e-04 |
| | $\tau_2$ | $\delta$ | 0.0254 | 9.8680e-05 |
| | | $\omega$ | 0.0065 | 1.5616e-05 |
| | $\tau_3$ | $\delta$ | 0.0421 | 1.6549e-04 |
| | | $\omega$ | 0.0064 | 1.5612e-05 |
| DoS-scenario 4 | $\tau_1$ | $\delta$ | 0.4605 | 0.0015 |
| | | $\omega$ | 0.1037 | 2.5907e-04 |
| | $\tau_2$ | $\delta$ | 0.0258 | 9.7460e-05 |
| | | $\omega$ | 0.0059 | 1.5288e-05 |
| | $\tau_3$ | $\delta$ | 0.0391 | 1.6378e-04 |
| | | $\omega$ | 0.0061 | 1.5604e-05 |

Table IV gives the state estimation indices of generator 1 based on the two algorithms under three DoS attack scenarios. From Table IV, it can be seen that: (1) For the RCKF, the values of $\tau_1$ are almost unchanged in three different attack scenarios. (2) For index $\tau_2$, under attack scenario 1, the values of the CKF are respectively 251 and 471 times that of those of the RCKF about the rotor angle and the rotor speed; under attack scenario 2, the values of the CKF are respectively 254 and 446 times that of those of the RCKF; under attack scenario 3, the values of the CKF are respectively 257 and 416 times that of those of the RCKF; under attack scenario 4, the values of the CKF are respectively 265 and 386 times that of those of the RCKF. (3) For the index $\tau_3$, the index values of the CKF are greater than those of the RCKF. In short, the effectiveness of the RCKF under DoS attacks is verified and the superiority of the RCKF compared with the CKF under DoS attacks is validated.

### D. NEW ENGLAND 16 MACHINE 68 BUS SYSTEM

This system consisting of 16 synchronous generators, 68 buses, and 86 lines is a well-known test system in the field of state estimation [1, 27]. Taken as generator 1 as an example, a three-phase short-circuit fault occurs at bus 6 at $t$=1s and then is removed at $t$=1.2s. The simulation lasts 10s and the detection threshold is here set to $B_J = 1.6$ using the same approach as that in the previous IEEE 9-bus test system.

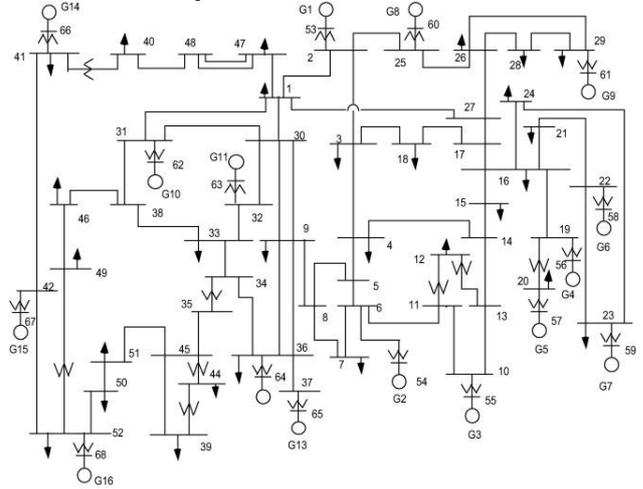

**FIGURE 20.** New England 16-machine-68-bus system

#### 1) SIMULATION RESULTS UNDER FDI ATTACKS

In order to facilitate comparative analysis, the settings of the FDI attack scenarios in this system are shown in Table V:

TABLE V
SCENARIOS of FDI ATTACKS

| Attack Scenarios | Error Vector of State Vector |
|---|---|
| FDI-scenario 1 | $c \sim N(0, \sigma^2), \sigma = 0.01$ |
| FDI-scenario 2 | $c \sim N(0, \sigma^2), \sigma = 0.1$ |
| FDI-scenario 3 | $c \sim N(0, \sigma^2), \sigma = 1$ |

In Table V, FDI-scenarios 1-3 respectively indicate that in three attack scenarios, the error vectors of the state vectors obey the Gaussian distribution with the mean 0 and the standard deviations 0.01, 0.1 and 1.

Figs. 21, 23 and 25 show that estimated results of the rotor angle in three FDI attack scenarios respectively; Figs. 22, 24 and 26 show that estimated results of the rotor speed under three attack scenarios respectively.

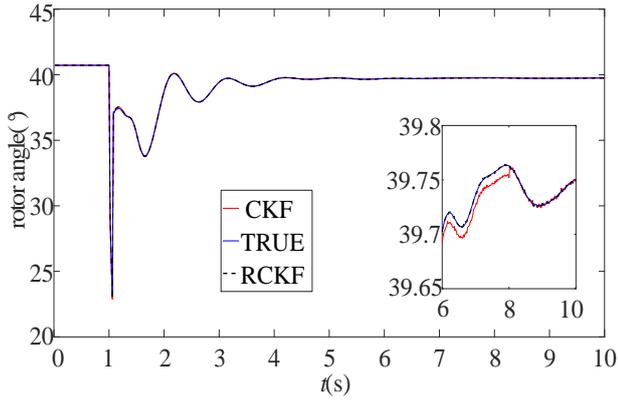

**FIGURE 21.** Rotor angle under FDI-scenario 1

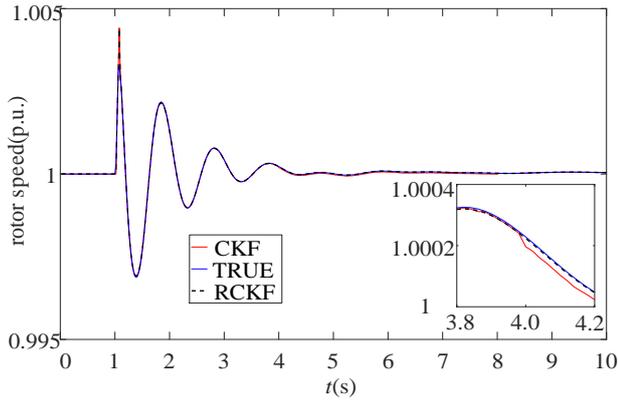

**FIGURE 22.** Rotor speed under FDI-scenario 1

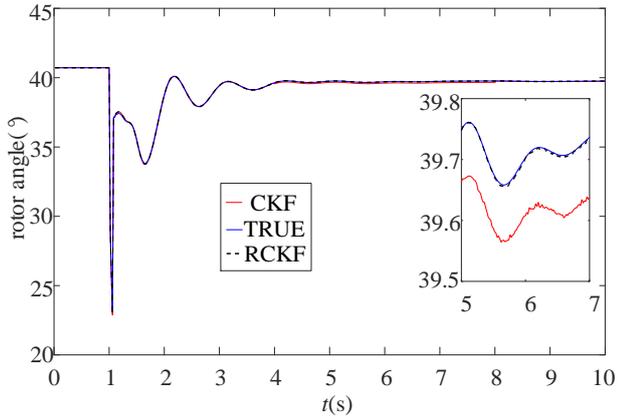

**FIGURE 23.** Rotor angle under FDI-scenario 2

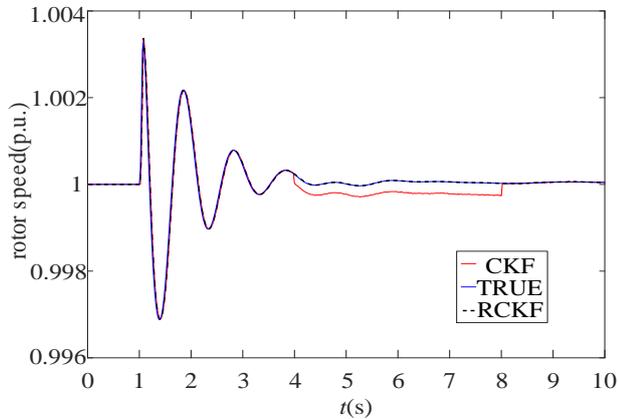

**FIGURE 24.** Rotor speed under FDI-scenario 2

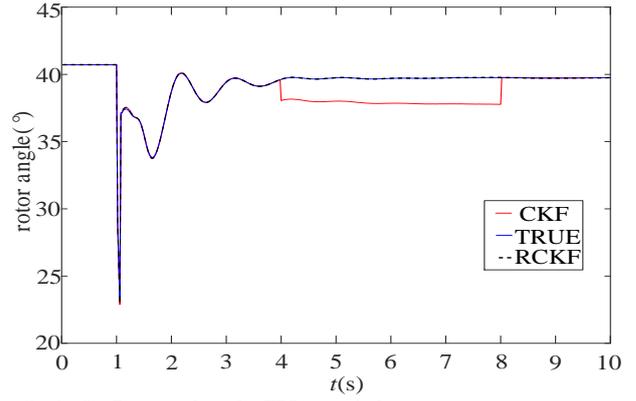

**FIGURE 25.** Rotor angle under FDI-scenario 3

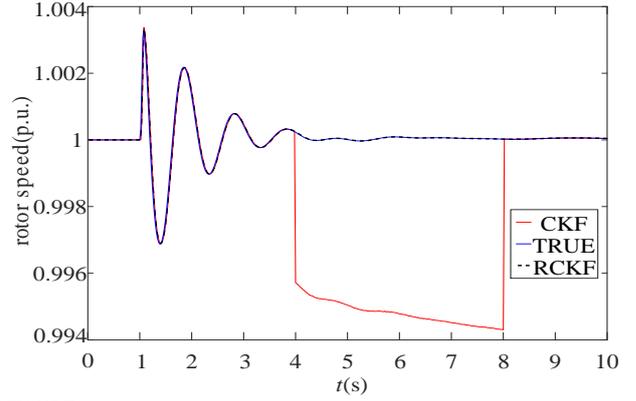

**FIGURE 26.** Rotor speed under FDI-scenario 3

From Figs. 21-26, it can be seen that the estimated values of the RCKF are significantly closer to the true values than those of the CKF in the attack period from $t$=4s to 8s. This verifies the effectiveness and applicability of the RCKF under three FDI attack scenarios for the larger system. And it also proves that the RCKF is better than the CKF in terms of the filtering ability.

The estimation indices of generator 1 are shown in Table VI.

Table VI
ESTIMATION INDICES OF GENERATOR 1

| Attack scenario | index | parameter | CKF | RCKF |
|---|---|---|---|---|
| FDI-scenario 1 | $\tau_1$ | $\delta$ | 5.0189e-04 | 4.9123e-04 |
| | | $\omega$ | 6.0381e-05 | 3.1985e-05 |
| | $\tau_2$ | $\delta$ | 7.5710e-04 | 5.5718e-04 |
| | | $\omega$ | 7.0899e-05 | 1.9432e-05 |
| | $\tau_3$ | $\delta$ | 2.7641e-05 | 4.0908e-05 |
| | | $\omega$ | 4.8011e-06 | 3.9398e-06 |
| FDI-scenario 2 | $\tau_1$ | $\delta$ | 0.0219 | 4.9588e-04 |
| | | $\omega$ | 0.0036 | 3.2311e-05 |
| | $\tau_2$ | $\delta$ | 0.0236 | 6.1624e-04 |
| | | $\omega$ | 0.0059 | 8.9818e-05 |
| | $\tau_3$ | $\delta$ | 0.0016 | 4.2342e-05 |
| | | $\omega$ | 2.5758e-04 | 3.9535e-06 |
| FDI-scenario 3 | $\tau_1$ | $\delta$ | 0.4335 | 4.9959e-04 |
| | | $\omega$ | 0.0743 | 3.2638e-05 |
| | $\tau_2$ | $\delta$ | 0.0440 | 9.9376e-05 |
| | | $\omega$ | 0.0068 | 5.1218e-06 |
| | $\tau_3$ | $\delta$ | 0.0321 | 4.4382e-05 |
| | | $\omega$ | 0.0053 | 3.9642e-06 |

Table V gives the state estimation indices of generator 1 based on the CKF and the RCKF under three FDI attack scenarios. From Table V, it can be observed that: (1) In the New England 16-machine 68-bus system, for the index $\tau_1$, the values of the rotor angle of the RCKF are respectively only increased by 0.9% and 0.7% with the diversification of attack scenarios, the values of the rotor speed of the RCKF don't change significantly. (2) In terms of the index $\tau_2$, the values of the CKF about the rotor angle and the rotor speed are 26.4% and 72.6% higher than those of the RCKF under attack scenario 1; under attack scenario 2, the values of the CKF are increased by 96.4% and 97.5% compared with those of the RCKF; under attack scenario 3, the values are increased by 98.8% and 98.9%. (3) For the index $\tau_3$, the values of the RCKF are always less than those of the CKF under three attack scenarios. These phenomena suggest that the filtering performance of the RCKF is still better than that of the CKF in the larger system.

2) SIMULATION RESULTS UNDER DOS ATTACKS

Similar to the FDI attacks, the settings of the DoS attack scenarios in this system are as same as those in the IEEE 9-bus system.

Figs. 27, 29, 31 and 33 show the estimated results of the rotor angle under different DoS attack scenarios. Moreover, Figs. 28, 30, 32 and 34 show the estimated results of the rotor speed.

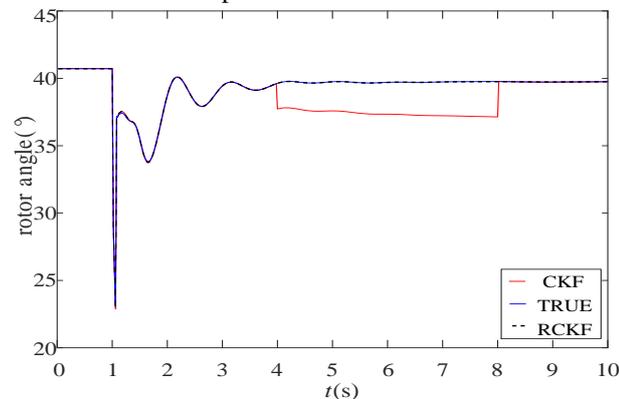

**FIGURE 27.** Rotor angle under DoS-scenario 1

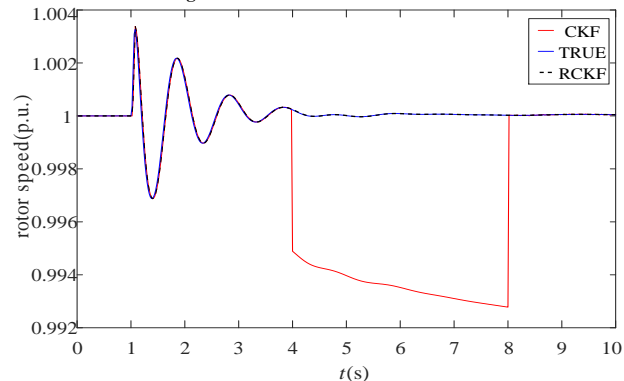

**FIGURE 28.** Rotor speed under DoS-scenario 1

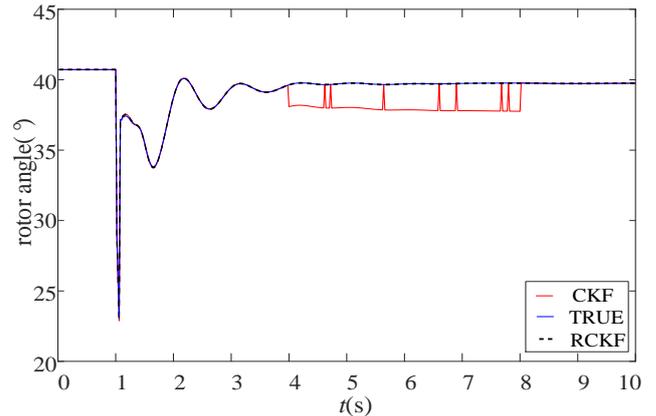

**FIGURE 29.** Rotor angle under DoS-scenario 2

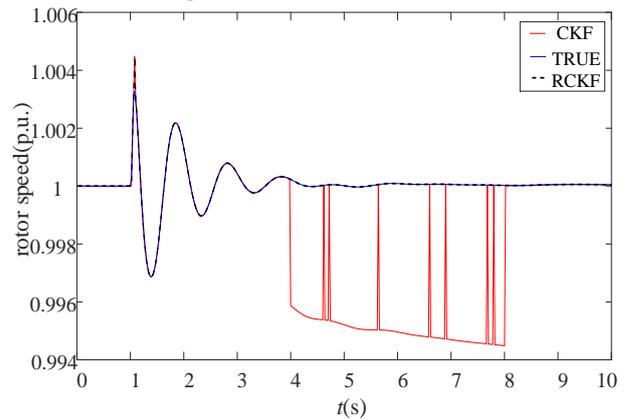

**FIGURE 30.** Rotor speed under DoS-scenario 2

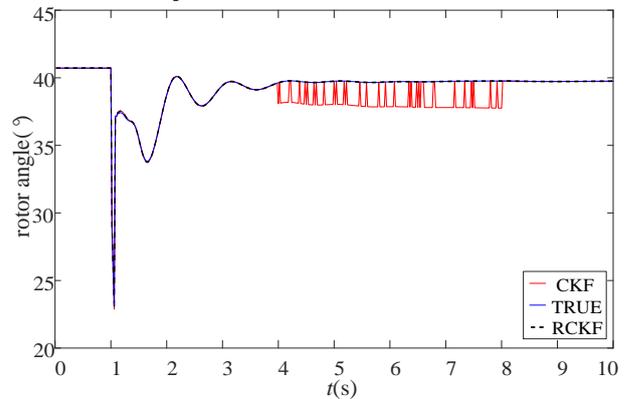

**FIGURE 31.** Rotor angle under DoS-scenario 3

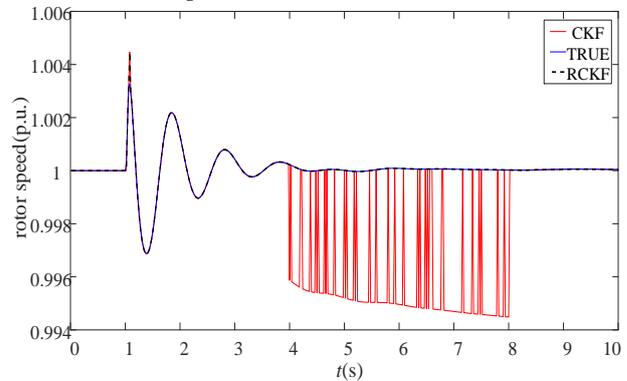

**FIGURE 32.** Rotor speed under DoS-scenario 3

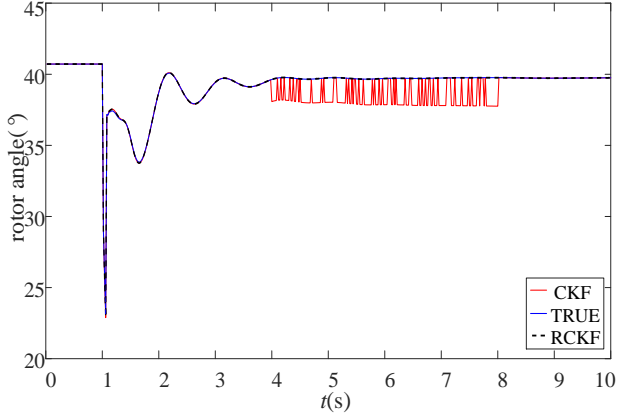

**FIGURE 33.** Rotor angle under DoS-scenario 4

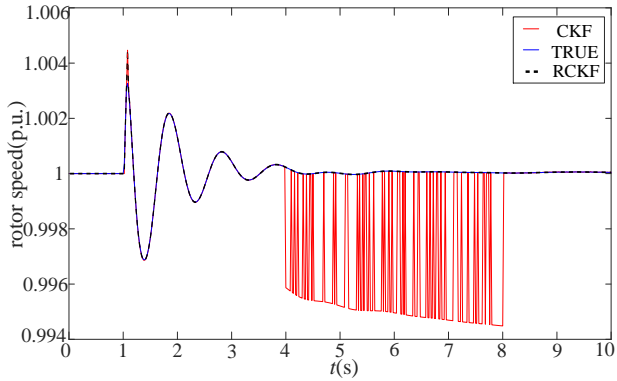

**FIGURE 34.** Rotor speed under DoS-scenario 4

From Figs. 27-34, it can be observed that the estimated values of the RCKF are still close to the true values with the different packet loss rates in this system. This verifies the effectiveness and adaptability of the RCKF under DoS attacks in the larger system. Moreover, it can be seen from these figures that the filtering ability of the RCKF is significantly superior to that of the CKF under DoS attacks.

The estimation indices of generator 1 are shown in Table VII.

Table VII
ESTIMATION INDICES OF GENERATOR 1

| Attack scenario | Index | Parameter | CKF | RCKF |
|---|---|---|---|---|
| DoS-scenario 1 | $\tau_1$ | $\delta$ | 0.5519 | 7.8748e-04 |
| | | $\omega$ | 0.0808 | 4.6549e-06 |
| | $\tau_2$ | $\delta$ | 0.0307 | 6.7476e-05 |
| | | $\omega$ | 0.0054 | 4.0097e-07 |
| | $\tau_3$ | $\delta$ | 0.0339 | 7.1068e-05 |
| | | $\omega$ | 0.0061 | 4.1030e-07 |
| DoS-scenario 2 | $\tau_1$ | $\delta$ | 0.4168 | 7.6748e-04 |
| | | $\omega$ | 0.0693 | 4.4549e-06 |
| | $\tau_2$ | $\delta$ | 0.0306 | 6.7233e-05 |
| | | $\omega$ | 0.0053 | 4.0980e-07 |
| | $\tau_3$ | $\delta$ | 0.0314 | 6.9065e-05 |
| | | $\omega$ | 0.0054 | 3.9477e-07 |
| DoS-scenario 3 | $\tau_1$ | $\delta$ | 0.3616 | 7.3863e-04 |
| | | $\omega$ | 0.0601 | 4.3292e-06 |
| | $\tau_2$ | $\delta$ | 0.0305 | 6.8115e-05 |
| | | $\omega$ | 0.0051 | 4.2346e-07 |
| | $\tau_3$ | $\delta$ | 0.0293 | 6.7146e-05 |
| | | $\omega$ | 0.0046 | 3.8729e-07 |
| DoS-scenario 4 | $\tau_1$ | $\delta$ | 0.3319 | 7.1382e-04 |
| | | $\omega$ | 0.0552 | 4.2560e-06 |
| | $\tau_2$ | $\delta$ | 0.0304 | 6.6464e-05 |
| | | $\omega$ | 0.0052 | 4.3880e-07 |
| | $\tau_3$ | $\delta$ | 0.0279 | 6.5003e-05 |
| | | $\omega$ | 0.0039 | 3.7557e-07 |

Table VI gives the state estimation indices of generator 1 based on these two algorithms under three DoS attack scenarios. From Table VI, it can be seen: (1) Regarding index $\tau_1$, the index values of the rotor angle and the rotor speed obtained by using the RCKF are basically unchanged under different attack scenarios, which validates the performance of the RCKF for the larger system under DoS attacks. (2) For index $\tau_2$, the index values of the rotor angle and the rotor speed obtained by the CKF are respectively 454 times and 13466 times higher than those of the RCKF under scenario 1; under scenario 2, the index values of the CKF are respectively increased by 454 times and 12932 times compared with those of the RCKF; under scenario 3, the index values of the CKF are respectively increased by 448 times and 12043 times compared with those of the RCKF; under scenario 4, the index values of the CKF increased by 456 times and 11622 times compared with those of the RCKF. (3) In terms of index $\tau_3$, the index values of the RCKF are always less than those of the CKF under the four scenarios. The results on the above indices $\tau_2$ and $\tau_3$ suggest that the RCKF significantly outperforms the CKF under DoS attacks.

From the test results on the above two systems, it can be seen that the filtering performance of the RCKF is superior to that of the CKF. The reason for this phenomenon is that the consequence of cyber attacks essentially is to introduce a large number of errors to the measurement data, and the RCKF can eliminate the errors while the CKF does not have this ability.

### E. COMPARISON OF COMPUTATIONAL EFFICIENCY

In order to reasonably evaluate the computational efficiency of the RCKF under cyber attacks, the calculation times using the CKF and the RCKF are shown in Table VIII.

Table VIII
CALCULATION TIME

| Attack | CKF(ms) | RCKF(ms) |
|---|---|---|
| FDI-scenario 1 | 0.123 | 0.137 |
| FDI-scenario 2 | 0.124 | 0.134 |
| FDI-scenario 3 | 0.127 | 0.132 |
| DoS-scenario 1 | 0.129 | 0.138 |
| DoS-scenario 2 | 0.128 | 0.139 |
| DoS-scenario 3 | 0.122 | 0.131 |
| DoS-scenario 4 | 0.124 | 0.136 |

Table VII shows that the RCKF can accurately estimate dynamic states of generators in real time. Since the RCKF uses the M estimation theory to eliminate measurement errors caused by cyber attacks, its calculation time is slightly longer than that of the CKF, but it is still within a reasonable range. Consequently, the computational times demonstrate that RCKF has the potential to perform DSE of generators or even power systems in real-world applications.

**VI. CONCLUSION**

This paper investigates the dynamic state estimation of generators under cyber attacks. First, attacks are modelled and thereby introduced into the DSE of generators; and then, the RCKF algorithm is adopted to estimate dynamic states of generators under cyber attacks with different degrees of sophistication. To the best of the authors' knowledge, this is the first study that investigates the DSE of generators under cyber attacks. Based on the test results on two IEEE test systems, the following conclusions can be drawn: (1) The RCKF is capable of effectively performing the DSE of generators in the presence of cyber attacks; (2) Furthermore, the filtering performance of the RCKF is significantly better than that of the CKF. (3) For a DSE algorithm such as the CKF that is not capable of addressing cyber attacks, the estimation performances of the algorithm may be seriously deteriorated under attacks.

In our future work, more types of cyber attacks will be introduced into the DSE of generators. In addition, the adopted method might be extended to address other power system state estimation issues.